# Investigation of Organic Nanocrystallic Films by the Method of Raman Spectroscopy

M. A. Korshunov[1]

*Kirensky Institute of Physics, Siberian Division, Russian Academy of Sciences, Krasnoyarsk, 660036 Russia*

**Abstract.** Investigation of organic molecular films by a method Raman of a spectroscopy is carried out. It is shown, that values of frequencies of lines in a spectrum of the lattice oscillations of the molecular films are depressed in comparison with values of frequencies of similar lines in a monocrystal. Intensity of padding lines will increase in a spectrum of films. Allocation of an impurity in a film of an organic mixed crystal nonuniform. Examinations are lead by the example of a p-dibromobenzene and the mixed crystal a p-dichlorobenzene / p-dibromobenzene.

In the molecular electronics the increasing application is discovered with organic molecular crystals (at making chips, and in devices of storage).

Thus, nanotechnologies are used. However, diminution of the sizes of crystal grains below some threshold quantity can result in to the considerable change of properties. For nuclear structures, this threshold makes about 100 nanometers. Change of physical properties because of change of the sizes of crystal grains for the structures constructed from organic molecules, apparently, should be observed at their greater size, than for the structures constructed from atoms.

Besides presence in structure of flaws (vacancies causing a diffusion, etc.) or impurities as can affect their properties and quality. For example, the molecule with the changed parameters can migrate and change noted information. As it is necessary to know, how located, impurities in received structure (it is uniformly or is not uniformly). Therefore the increasing value is gained with methods of non-destructive quality assurance used is model also examination of received structure of a Nanocrystallic material.

For study of these problems *investigation* of a Nanocrystallic film of organic matters by a method Raman of a spectroscopy has been carried out. This method has been chosen, as spectrums of the lattice oscillations are rather sensitive to structural change and a composition of a crystal.

As objects of research molecular crystals a p-dibromobenzene and a solid solution p-dichlorobenzene /p-dibromobenzene, which are good model, objects have been chosen. These substances are investigated by various methods. They crystallize in space group $P2_1/a$ with two molecules in a unit cell [1]. The p-dibromobenzene and p-dichlorobenzene form mix-crystals at any concentrations of builders. The sample has been prepared as follows. On a glass plate (integumentary glass) was evaporated film investigated substance. It has formed a film thickness from ~ 0.7 up to 5 μ. From above the received film has been covered with other integumentary glass as the substance flashes, reducing thickness of a film. After that Raman spectrums as the lattice, and the intramolecular oscillations up to 300 $cm^{-1}$ have been received.

---

[1] E-mail: mkor@iph.krasn.ru

In figure the unpolarized spectrums of the lattice oscillations received for a film of a p-dibromobenzene (а) and his monocrystal (б) are given. In a spectrum of the lattice oscillations it should be observed six intensive lines of the molecules stipulated by rotary huntings around of moments of inertia.

The magnification of intensity of padding lines can be stipulated by that per acre a film is necessary more flaws, than on a unit volume of a monocrystal, and also influence of boundaries of a film and more legible display of the surface oscillations. Apparently, influence of boundaries of a film as affects and diminution of values of frequencies of some lines.

And a series of padding lines of small intensity of the selection rules stipulated by infringement because of presence of flaws. As it visual, in a spectrum of a film padding lines have major intensity, than in a spectrum of a monocrystal. Except for that value of frequencies of similar most intensive lines in a spectrum of a film have smaller value, than in a spectrum of a monocrystal. For three most legiblly apparent lines of value of frequencies for a film and a monocrystal are given in the table.

Table. Values of frequencies (ν) for a monocrystal and a film.

| Frequency | $\nu_1$ | $\nu_5$ | $\nu_6$ |
|---|---|---|---|
| Monocrystal | 20.0 | 93.0 | 97.0 |
| Film | 18.3 | 88.0 | 93.0 |

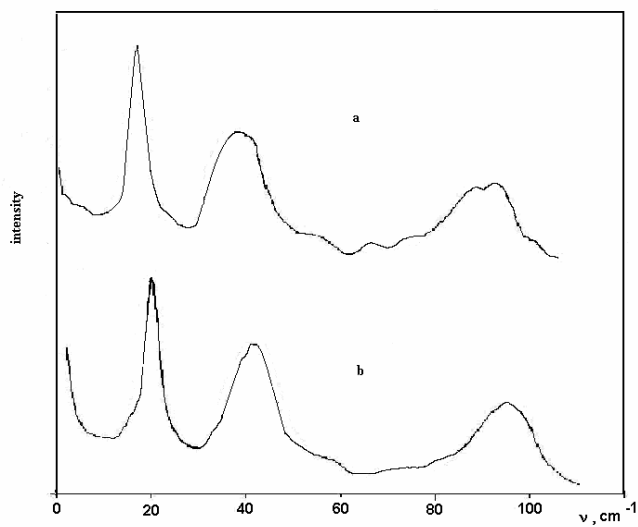

Fig. A spectrum of the lattice oscillations of a p-dibromobenzene for a film (a) and a monocrystal (b).

For definition of allocation of an impurity (p-dibromobenzene) in a film of the mixed crystal a p-dichlorobenzene/p-dibromobenzene with concentration of a p-dibromobenzene ~ 20 mol. % shootings Raman spectrums are carried out. Spectrums are received in various points of a film. Definition of concentration of an impurity was carried out on a method offered in work [2]. It is found, that allocation of an impurity in a plane of a film is nonuniform. But in due course it can vary because of a diffusion. It is possible to assume, as allocation of vacancies in a film as is chaotic, as well as allocation of an impurity.

Thus, it is shown, that at investigation of Nanocrystallic films of organic crystals it is possible to use a method of a Raman effect, for investigation of allocation of an impurity in a film, and study of influence of the sizes of crystals on dynamics of lattices. Besides the organic molecular of a film can be viewed as a two dimensional crystal in comparison with a three-dimensional monocrystal that allows to carry out comparison of calculations of dynamics of a lattice for two dimensional and three-dimensional objects.